\documentclass[prb,preprint,eqsecnum,aps]{revtex4}
\usepackage{graphicx}
\usepackage{epstopdf}
\newcommand{\ddt}{{\Delta\tau}}
\newcommand{\dt}{{\epsilon}}
\newcommand{\e}{{\rm e}}
\newcommand{\al}{\alpha}

\newcommand{\ep}{\epsilon}

\newcommand{\ga}{\gamma}
\newcommand{\ka}{\kappa}

\newcommand{\tka}{\tilde\kappa}
\newcommand{\tmu}{\tilde\mu}
\newcommand{\za}{\zeta}

\newcommand{\bea}{\begin{eqnarray}}
\newcommand{\eea}{\end{eqnarray}}
\newcommand{\be}{\begin{equation}}
\newcommand{\ee}{\end{equation}}
\newcommand{\ba}{\begin{eqnarray}}
\newcommand{\ea}{\end{eqnarray}}

\newcommand{\nn}{\nonumber}
\newcommand{\la}{\label} 
\newcommand{\w}{\omega} 
\newcommand{\hT}{\hat{T}}
\newcommand{\hG}{\hat{G}}
\newcommand{\hV}{\hat{V}}

\newcommand{\hH}{\hat{H}}

\def\t1{e_{_T}}
\def\v1{e_{_V}}

\begin{document}
\title{Anatomy of Path Integral Monte Carlo: algebraic derivation of the harmonic oscillator's universal 
	discrete imaginary-time propagator and its sequential optimization}

\author{Siu A. Chin}
\email{chin@physics.tamu.edu.}

\affiliation{Department of Physics and Astronomy, 
	Texas A\&M University, College Station, TX 77843, USA}

\begin{abstract}

The direct integration of the harmonic oscillator path integral obscures the
fundamental structure of its discrete, imaginary time propagator (density matrix).
This work, by first proving an operator identity for contracting two free propagators into
one in the presence of interaction, derives the discrete propagator
by simple algebra without doing any integration. This discrete propagator is $universal$, having the same
two hyperbolic coefficient functions for all short-time propagators. 
Individual short-time propagator only modifies the coefficient function's
argument, its $portal$ parameter, whose convergent order is the same as the thermodynamic energy. 
Moreover, the thermodynamic energy can be given in a closed form for any short-time propagator.
Since the portal parameter can be systematically optimized by matching the expansion
of the product of the two coefficients, any short-time propagator can be optimized $sequentially$, 
order by order, by matching the product coefficient's expansion alone, without computing the energy.
Previous empirical findings on the convergence of fourth and sixth-order propagators can now
be understood analytically. An eight-order convergent short-time propagator is also derived.

\end{abstract}


\maketitle

\section {Introduction}

Current advances in Path Integral Monte Carlo (PIMC) continue to be the use of parametrized fourth-order short-time propagators to optimize PIMC's convergence at the fewest number of 
beads \cite{jan01,sak09,chin15,kam16,lin18,wan22}. 
This is especially important for ameliorating the sign problem in
fermion systems\cite{chin15}. Since the harmonic oscillator's discrete imaginary 
time path integral can be computed analytically\cite{kam16,wan22,sch81}, 
it has been widely used as a testing ground for such optimization. However, even for the
harmonic oscillator, such multi-parameter optimization is non-trivial and has been done only 
by trial and error in PIMC\cite{sak09}, or by elaborate theoretical computations\cite{wan22}.

In this work, we solve for, and optimize, the discrete harmonic oscillator path integral by a
completely novel approach, unique to the harmonic oscillator. The method therefore may not be
generalizable to more realistic interactions. However, what has achieved here for the harmonic oscillator,
will at least suggest what might be possible for future PIMC.

The conventional method of solving for the harmonic oscillator's discrete path integral is by 
direct integration\cite{kam16,wan22,sch81}. Manipulating the resulting 
tridiagonal matrix is akin to manipulating a black box, with no easy comprehension 
of what's inside. In this work, we completely abandon this opaque approach. Instead,  
we first prove in Sect.\ref{oc}, an operator identity
for contracting two free propagators into one in the presence of the harmonic interaction. 
Once proven, this identity eliminates the need for doing any integral.
By applying this identity to left-right symmetric short-time propagators in Sect.\ref{mpro}, this work
derives the discrete, imaginary-time harmonic oscillator propagator by simple algebra.
This discrete propagator has a {\it universal} structure, having the 
same functional form, the same two hyperbolic coefficient functions, for all short-time propagators. 
Different short-time propagator only changes the {\it portal} parameter, the argument of the 
hyperbolic functions.

This universal discrete propagator now opens up new ways of optimizing any
short-time propagator due to two results in Sect.\ref{corder}. First,
the convergence of the portal parameter can be determined by matching the 
series expansion of the product of the two coefficients to the exact result.
Second, the convergent order of the portal parameter can be shown to be the same 
as the thermodynamic energy.
Therefore, the parameters in the short-time propagator, through the portal parameter,
can be optimized order-by-order, by matching the series expansion of the
product coefficient, without computing the energy. The advantage here is that
one does not need to optimize all short-time propagator parameters {\it simultaneously} in minimizing
the energy. Each parameter can be optimized {\it sequentially}, by matching the order conditions
one by one. 

By applying this optimization procedure to fourth and sixth-order short-time-propagators 
in Sect.\ref{opfour}, one can now derive analytically 
the optimal parameter values empirically found by Sakkos {\it et al.}\cite{sak09} 
and reproduce those computed theoretically by Wang {\it et al.}\cite{wan22}. In order
to compare published results for these high-order propagators, the thermodynamic
energy is derived in a closed form for any short-time propagator.  
Finally, in Sect.\ref{eighto}, the same optimization procedure can be exploited to yield
a new eight-order algorithm for solving the harmonic oscillator. 
 
In Sect.\ref{pi}, 
we give a brief review of PIMC, emphasizing its
the operator formulation.
Conclusions and future directions are stated in Sect.\ref{con}.

\section {Discrete Path integral}
\la{pi}

In suitable units, the one-dimensional imaginary time Schr\"odinger equation
can always be written as 
\be
-\frac{\partial\psi(x,\tau)}{\partial \tau}=(\hat T+\hat V)\psi(x,\tau)=
\left(-\frac12\frac{\partial^2}{\partial { x}^2}
+V(x)\right)\psi(x,\tau),
\ee
with dimensionless spatial variable $x$ and imaginary time $\tau$($\propto\beta\!=\!1/k_BT$). 
The solution in operator form is just
\be
|\psi(\tau)\rangle=\hG(\tau)|\psi(0)\rangle,
\ee
where the imaginary time evoluation operator $\hG(\tau)$ is given by
\be
\hG(\tau)=\e^{-\tau(\hat T+\hat V)}.
\ee
The matrix element of $\hG(\tau)$ is the propagator, or density matrix:
\be
G(x^\prime,x,\tau)=\langle x^\prime|\hG(\tau)|x\rangle
=\sum_n\psi_n^*(x^\prime)\psi_n(x)\e^{-\tau E_n}.
\ee
In PIMC, one is
interested in extracting the square of the ground state wave function $\psi^2_0(x)$ and energy $E_0$ 
from the long time (or low temperature) limit of the imaginary time propagator:
\be
\lim_{\tau\rightarrow\infty}G(x,x,\tau)\longrightarrow\psi_0^2(x)\e^{-\tau E_0}+ \cdots.
\ee

Since the matrix element of  $\hG(\tau)$ is generally not known analytically, it is usually approximated by
a product of $N$ short-time operators at imaginary time-step $\dt=\tau/N$,
\be
\hG_N=[\hG_1(\dt)]^N,
\la{gn}
\ee
where $\hG_1(\dt)$ is a short-time approximation of $\hG(\dt)$
\ba
\hG_1(\dt)&\approx&\e^{-\dt(\hat T+\hat V)},
\ea
of which the simplest is the second-order {\it primitive approximation} (PA) 
\ba
\hG_1(\dt)&=\e^{-\frac12\dt\hat V}\e^{-\dt\hat T}\e^{-\frac12\dt\hat V},
\la{pao}
\ea
with matrix element
\ba
&&G_1(x_1,x_2,\dt)=\langle x_1|\hG_1(\dt)|x_2\rangle\nn\\
&&\qquad =\frac1{\sqrt{2\pi \dt}}\e^{-\frac12\dt V(x_1)}
\e^{-\frac1{2\dt}(x_1-x_2)^2}
\e^{-\frac12\dt V(x_2)}.
\la{pa}
\ea
The matrix element of (\ref{gn}) is then,
\ba
G_{N}(x^\prime,x,\tau)&=&\langle x^\prime|[\hG_1(\dt)]^N|x\rangle\nn\\
&=&\int_{-\infty}^{\infty} dx_1\cdots dx_{N-1}\,
G_1(x^\prime,x_1,\dt)G_1(x_1,x_2,\dt)\cdots G_1(x_{N-1},x,\dt),
\la{mb}
\ea
which is the conventional path integral for the discrete propagator
at $\tau=N\dt$.

In this work, because of the special nature of the harmonic oscillator, it is far more
transparent to work directly at the simpler operator level (\ref{gn}) than at the convoluted
path integral level (\ref{mb}). Moreover, this operator approach can easily accommodate
 higher-order short-time operator $\hG_1$ more complex
than the primitive approximation (\ref{pao}).  


\section {Operator contractions}
\la{oc}

In terms of dimensionless length $x=r/\sqrt{\hbar/m\w}$, energy $E={\cal E}/\hbar\w$
and imaginary time $\tau=\hbar\w/k_BT$, the dimensionless harmonic potential operator is $\hV(x)=x^2/2$.
In this case, the Gaussion integral below can be computed to give
\ba
\langle x_1|\e^{ -a\hT}\e^{ -b\hV}\e^{-c\hT}|x_2\rangle
&=&\frac1{\sqrt{2\pi a}}\frac1{\sqrt{2\pi c}}
\int_{-\infty}^{\infty} dx\, 
\e^{ -\frac1{2a}(x_1-x)^2-b\frac12 x^2-\frac1{2c}(x-x_2)^2}\la{gi1}\\
&=&\frac1{\sqrt{2\pi \ka}}\e^{-\nu\frac12 x_1^2}
\e^{-\frac1{2\ka}(x_1-x_2)^2}\e^{-\mu\frac12 x_2^2}\la{gmat}\\
&=&\langle x_1|\e^{ -\nu\hV}\e^{-\ka\hT}\e^{ -\mu\hV}|x_2\rangle
\la{gi2}
\ea
where
\be
 \ka=a+abc+c,\quad\nu=\frac{bc}{\ka},\quad\mu=\frac{ba}{\ka}.
\la{kauv}
\ee
This means that (\ref{gi2}) is actually an operator identity
\be
\e^{ -a\hT}\e^{ -b\hV}\e^{-c\hT}=\e^{ -\nu\hV}\e^{-\ka\hT}\e^{ -\mu\hV},
\la{opid}
\ee
where two $\hT$ operators, in the presence of interaction $\hV$, have been {\it contracted} into one. 
Therefore, any numbers of operators of $\hT$ and $\hV$ for the harmonic oscillator can be contracted down
to a single $\hT$ operator form, as in the RHS of (\ref{opid}), with obvious matrix element (\ref{gmat}). 

Applying (\ref{opid}) twice to the following three-operator case gives
\ba
\e^{ -a\hT}\e^{ -b\hV}\e^{-c\hT}\e^{ -b\hV}\e^{ -a\hT}
&=&\e^{ -\nu\hV}\e^{-\ka\hT}\e^{ -\mu\hV}\e^{ -b\hV}\e^{ -a\hT},\la{fapp}\\
&=&\e^{ -\nu\hV}\e^{ -\nu'\hV}\e^{-\ka'\hT}\e^{ -\mu'\hV},\nn\\
&=&\e^{ -\mu'\hV}\e^{-\ka'\hT}\e^{ -\mu'\hV},
\la{threet}
\ea
where the second application at (\ref{fapp}) with $a'=\ka$, $b'=\mu+b$ and $c'=a$ yields
\ba
\ka'&=&\ka+\ka(\mu+b)a+a=(\ka+a)+ba(\ka +a)\nn\\
&=&(1+ab)(\ka+a)=(1+ab)(2a+abc+c)
\ea
\be
\mu'=\frac{(\mu+b)\ka}{\ka'}
=\frac{b(a+\ka)}{(1+ab)(\ka+a)}=\frac{b}{(1+ab)}
\ee
\ba
\nu+\nu'&=&\frac{bc}{\ka}+\frac{(\mu+b)a}{\ka'}
=\frac{bc}{\ka}+\frac{ab}{\ka'}\left(\frac{a+\ka}{\ka}\right)=\mu'.
\ea
Note that the product
\be
\ka'\mu'=b(2a+abc+c),
\la{poly}
\ee
is only a polynomial in the original coefficients.

It follows that any palindromic, left-right symmetric approximate short-time operator of the form 
\be
\hG_1(\dt)=\prod_{i}\e^{ -a_i\dt\hT}\e^{ -b_i\dt\hV},
\ee 
can be contracted down to a  single $\hT$-operator, palindromic form
\be
\hG_1(\dt)=\e^{ -\mu_1\hV}\e^{ -\ka_1\hT}\e^{ -\mu_1\hV},
\la{sho}
\ee 
where $\mu_1$ and $\ka_1$ are functions of $a_i$, $b_i$ and $\dt$. Once
contracted to a single $\hT$ operator form, its matrix element is then easily given
by (\ref{gmat}). Therefore we will refer to the short-time operator interchangeably
as the short-time propagator.

For the PA short-time propagator,
one has simply
\be 
\ka_1=\dt\quad{\rm and}\quad \mu_1=\frac12 \dt.
\la{pai}
\ee
Other higher order short-time operators will be examined in Sects.\ref{corder} and \ref{opfour}

 
\section {The universal discrete propagators}
\la{mpro}

Starting with a palindromic short-time propagator $\hG_1$, any product $\hG_1^N$ must also
be palindromic, contractable down to a single $\hT$-operator form. 
This is obvious because if $N$ is even then operators on both sides are equal. If $N$ is odd, then
there is central palindromic short-time operator with equal number of $\hG_1$ on both sides.
Therefore,
\ba
\hG_{m+n}&=&[\hG_1(\dt)]^m[\hG_1(\dt)]^n\la{mn}\\
\e^{ -\mu_{m+n}\hV}\e^{ -\ka_{m+n}\hT}\e^{ -\mu_{m+n}\hV}
&=&\e^{ -\mu_m\hV}\e^{ -\ka_m\hT}\e^{ -\mu_m\hV}\e^{ -\mu_n\hV}\e^{ -\ka_n\hT}\e^{ -\mu_n\hV}\nn\\
&=&\e^{ -\mu_m\hV}[\e^{ -\ka_m\hT}\e^{ -(\mu_m+\mu_n)\hV}\e^{ -\ka_n\hT}]\e^{ -\mu_n\hV}\nn\\
&=&\e^{ -\mu_m\hV}\e^{ -\nu\hV}\e^{-\ka\hT}\e^{ -\mu\hV}\e^{ -\mu_n\hV}
\la{gmn}
\ea
where, by the contraction identity (\ref{opid})
\ba
\ka&=&\ka_m+\ka_n+\ka_m(\mu_m+\mu_n)\ka_n=\ka_{m+n},\la{kmn}\\
\nu&=&\frac{(\mu_m+\mu_n)\ka_n}{\ka_{m+n}},\nn\\
\mu&=&\frac{(\mu_m+\mu_n)\ka_m}{\ka_{m+n}}.
\ea
If one now defines an equally fundamental, {\it product variable} of $\ka_n$ and $\mu_n$ via
\be
\za_n=1+\ka_n\mu_n,
\la{zf}
\ee
with $\za_1=1+\ka_1\mu_1$ known from $\hG_1$, then (\ref{kmn}) reads
\ba
\ka_{m+n}
&=&\ka_m+\ka_n+\ka_n(\za_m-1)+\ka_m(\za_n-1)\nn\\
&=&\ka_n\za_m+\ka_m\za_n.
\la{kadd}
\ea
Since (\ref{gmn}) is palindromic, one must have
\ba
\mu_{m+n}=\mu_m+\frac{\ka_n(\mu_m+\mu_n)}{\ka_n\za_m+\ka_m\za_n}
&=&\mu_n+\frac{\ka_m(\mu_m+\mu_n)}{\ka_n\za_m+\ka_m\za_n}
\la{mad2}
\ea
The last equality can be true only if
\be
\mu_m(\ka_n\za_m+\ka_m\za_n)+\ka_n(\mu_m+\mu_n)=\mu_n(\ka_n\za_m+\ka_m\za_n)+\ka_m(\mu_m+\mu_n).
\ee
Substituting in $\ka_n\mu_n=\za_n-1$ gives,
\ba
&&\mu_m\ka_n\za_m+(\za_m-1)\za_n+\ka_n\mu_m+(\za_n-1)\nn\\
    &&\qquad\qquad =(\za_n-1)\za_m+\mu_n\ka_m\za_n+(\za_m-1)+\ka_m\mu_n,
\ea
which simplifies to
\be
\mu_m\ka_n\za_m+\ka_n\mu_m=\mu_n\ka_m\za_n+\ka_m\mu_n.
\ee
Since $\mu_k=(\za_k-1)/\ka_k$, this requires that
\ba
\frac{\ka_n(\za^2_m-1)}{\ka_m}&=&\frac{\ka_m(\za^2_n-1)}{\ka_n},\nn\\
\Rightarrow\quad \frac{(\za^2_m-1)}{\ka^2_m}&=&\frac{(\za^2_n-1)}{\ka^2_n},
\ea
implying that $(\za_k^2-1)/\ka_k^2$ must be a constant, say $\ga^2$, for all $k\ge 1$: 
\be
\frac{\za_k^2-1}{\ka_k^2}=\frac{\za_1^2-1}{\ka_1^2}\equiv\ga^2.
\la{const}
\ee
This constant $\ga$ is determined solely by the short-time operator $\hG_1$. 

Given (\ref{const}), the coefficient function $\mu_{m+n}$ is then given  by
\be
\mu_{m+n}=\frac{(\ga\ka_m)(\ga\ka_n)+\za_m\za_n-1}{\ka_n\za_m+\ka_m\za_n}.
\la{madd}
\ee
This means that if one defines a new set of variables
\be
\tilde\ka_k=\ga\ka_k,\qquad \tilde\mu_k=\mu_k/\ga, 
\la{new}
\ee
rewrite (\ref{kadd}) as
\be
\tka_{m+n}=\tka_n\za_m+\tka_m\za_n,
\la{tadd}
\ee
then (\ref{madd}) can be written simply as
\be
\tmu_{m+n}=\frac{\za_{m+n}-1}{\tka_{m+n}},
\la{hmu}
\ee
with
\be
\za_{m+n}=\tka_m\tka_n+\za_m\za_n.
\la{zadd}
\ee
In terms of the new variables (\ref{new}), $\za_k$ is unchanged, given by 
$\za_k=1+\tka_k \tmu_k$, with (\ref{const}) now takes the form
\be
\za_k^2-\tka_k^2=1.
\la{sq}
\ee

Eqs.(\ref{tadd}) and (\ref{zadd}) define the addition of
$\tka_k$ and $\za_k$ identical to the addition of hyerbolic sine and cosine functions.
Since the addition begins with $\tka_1$ and $\za_1$, if one defines a new variable $u$ related to $\dt$
such that
\be
\za_1(\dt)=\cosh(u),
\la{xdef}
\ee
then (\ref{sq}) implies that
\be
\tka_1(\dt)=\sinh(u),
\la{kdef}
\ee
and the addition formulas {\it instantaneously} yield, without doing any integration,
\ba
&&\za_N=\cosh(Nu),\la{hza} \\
&&\tka_N=\sinh(Nu),\la{kka} \\
&&\tmu_N=\frac{\cosh(Nu)-1}{\sinh(Nu)}=\tanh(Nu/2).
\la{mma}
\ea
The right hand sides above are the {\it universal} coefficient functions of the discrete propagator with
$N$ beads, or $N$ short-time propagators. They are the same for all short-time operators. 
All short-time operators enter into this universal propagator through the {\it portal} parameter 
$u$ via (\ref{xdef}),
\be
u=\log\left(\za_1+\sqrt{\za_1^2-1}\right).
\la{udef}
\ee
In term of $u$, (\ref{kdef}) can be used to define
\be
\ga=\sinh(u)/\ka_1,
\ee
making (\ref{kka}) explicit as
\be
\ka_N=\frac1{\ga}\sinh(Nu)=\frac{\ka_1}{\sinh u}\sinh(Nu).
\la{knx}
\ee
(We note that (\ref{hza}) and (\ref{knx}) can be expressed directly in terms of $\za_1$,
via Chebyshev polynomials $T_N(\za_1)$ and $U_{N-1}(\za_1)$ via (\ref{udef}) as
\ba
\za_N&=&\frac12 \left[\left(\za_1+\sqrt{\za_1^2-1}\right )^N
+ \left(\za_1-\sqrt{\za_1^2-1}\right)^N\right]=T_N(\za_1), \la{zn}
\ea
and 
\ba
\ka_N&=&\frac{\ka_1}{\sqrt{\za_1^2-1}}\sinh(Nu),\nn\\
&=&\frac{\ka_1}{2\sqrt{\za_1^2-1}}\left[\left(\za_1+\sqrt{\za_1^2-1}\right )^N
-\left(\za_1-\sqrt{\za_1^2-1}\right)^N\right]=\ka_1 U_{N-1}(\za_1).
\la{kn}             
\ea
This is of interest for obtaining the analytical forms of $\za_N$ and $\ka_N$. For
numerical results, the use of the portal parameter forms (\ref{hza}) and (\ref{knx})
is more efficient.)

One can now compare the discrete propagator derived here with those already published.
For the ease of comparison, one can write the discrete propagator below in two forms:
\ba
G_N(x',x;\tau)&=&\langle x'|\e^{ -\mu_N\hV}\e^{ -\ka_N\hT}\e^{ -\mu_N\hV}|x\rangle\nn\\
&=&\frac1{\sqrt{2\pi \ka_N}}\exp\left [-\mu_N\frac12 {x'}^2
-\frac1{2\ka_N}(x'-x)^2-\mu_N\frac12 x^2\right ]\la{sf}\\
&=&\frac1{\sqrt{2\pi \ka_N}}\exp\left [-\frac{1}{2\ka_N}\left (\za_N(x^2+{x'}^2)-2xx'\right ) \right ].
\la{af}
\ea
The second form (\ref{af}) only uses the two fundamental hyperbolic functions $\za_N$ (\ref{hza})
and $\ka_N$ (\ref{knx}).
For the PA propagator, from (\ref{pai}),
\be
\za_1=1+\frac12\ep^2,\qquad u=\cosh^{-1}(1+\frac12\ep^2),
\la{paz}
\ee
and one immediately obtains, by mere inspection,
\be
\za_N=\cosh(Nu),\qquad \ka_N=\frac{\ep}{\sinh u}\sinh(Nu).
\la{knpa}
\ee
The above agrees with Kamibayashi and Miura's\cite{kam16} result, with their
$\beta=N\Delta\tau$ equivalent to $\tau=N\ep$ and their $\theta$ in place of $u$.

In the original tridiagonal matrix calculation of Schweizer {\it et al.}\cite{sch81}, $\ka_N$
is given as (after setting $\hbar=m=\omega_0=1$)
\be
\ka_N=\ep\left (\frac{f}{f^2-1}\right )\left (\frac{f^{2N}-1}{f^N}\right)
\la{knold}
\ee
where the rather obscure $f$ turns out to be
\be
f=1+\frac12\dt^2+\frac12 \dt\sqrt{\dt^2+4}=\za_1+\sqrt{\za_1^2-1}=\e^u
\ee
and therefore (\ref{knold}) is precisely (\ref{knpa}).

Fourth-order propagators will be considered in Sect.\ref{opfour}.


\section {Convergent order of short-time operators}
\la{corder}

For the PA propagator, as noted previously $\za_1=1+\dt^2/2$.
In the {\it convergent limit} of $\dt=\tau/N\rightarrow 0$ while keeping $\tau$ fixed, 
\ba
u&=&\cosh^{-1}(\za_1)\la{ucosh}\\
  &\rightarrow&\dt-\frac1{24}\dt^3+\cdots,\nn\\
Nu&\rightarrow&N\dt=\tau,
\ea
one obtains the coefficient (\ref{hza})-(\ref{kka}):
\ba
&&\za(\tau)=\cosh(\tau),\la{eza} \\
&&\ka(\tau)=\sinh(\tau),\la{eka}
\la{ema}
\ea
reproducing the exact propagator\cite{fey72}
\be
G(x^\prime,x,\tau)
=\frac1{\sqrt{2\pi \ka(\tau)}}\exp\left [-\frac{1}{2\ka(\tau)}\left (\za(\tau)(x^2+{x'}^2)-2xx'\right )\right ],
\la{exactp}
\ee
since $\ga\rightarrow 1$ as $\dt\rightarrow 0$. 

While (\ref{exactp}) verifies the correct convergence of our discrete propagator, this exact result is
not the focus of this work. Instead, we are interested in optimizing
$\hG_N$ so that it can approach this result at a minimum value of $N$. 

Clearly the minimum value is $N=1$ if 
\be
\za_1(\dt)=\cosh(\dt),
\la{zacos}
\ee
forcing $u=\dt$ in (\ref{ucosh}), so that $\hG_1$ is the exact propagator.
If $\hG_1$ is not exact, then its coefficient $\za_1(\dt)$ must match the expansion 
\be
\cosh(\dt)=1+\frac1{2!}\ep^2+\frac1{4!}\ep^4+\frac1{6!}\ep^6+\cdots
\la{chexp}
\ee
as closely as possible, order by order in $\ep$.
By comparing PA's $\za_1$ from (\ref{paz}) to the above expansion
one sees that it is a second-order algorithm because it matches the above expansion 
to second order in $\dt$. Therefore, {\it the order of the propagator is just the correct 
expansion order of its product coefficient function $\za_1(\dt)$}.
This is an {\it intrinsic} characterization of the convergence order of any short-time propagator
for the harmonic oscillator, with no reference to any calculations {\it extrinsic} to the algorithm. 

However, by the following calculation of the thermodynamic energy,
this intrinsic characterization is the same as the conventional definition 
that the order of the propagator is the order of its energy error.

First, we note that if a short-time propagator's $\za_1$ is correct up to order $2n$, 
\be
\za_1=1+\frac1{2!}\ep^2+\cdots +\frac1{(2n)!}\ep^{2n}+\frac{\delta_{2n+2}}{(2n+2)!}\ep^{2n+2}
\ee
but has error at order $2n+2$ such that $\delta_{2n+2}< 1$,
then the matching condition (\ref{xdef}) would force 
the portal parameter to converge as (note the minus sign)
\be
u=\ep(1-C\ep^{2n}+\cdots)
\la{uep}
\ee
with $C$ determined by
\ba
\sum_{k=0}^n\frac{\ep^{2k}}{(2k)!}+\frac{\delta_{2n+2}}{(2n+2)!}\ep^{2n+2}
&=&\sum_{k=0}^{\infty}\frac{\ep^{2k}}{(2k)!}(1-C\ep^{2n})^{2k}
=\sum_{k=0}^{\infty}\frac{\ep^{2k}}{(2k)!}(1-2k C\ep^{2n}+\cdots)\nn\\
\frac{\delta_{2n+2}}{(2n+2)!}\ep^{2n+2}&=&\frac1{(2n+2)!}\ep^{2n+2}-C\ep^{2n+2}+\cdots
\ea
to be
\be
C=\frac{1-\delta_{2n+2}}{(2n+2)!}.
\la{ccof}
\ee
so that (\ref{uep}) reads
\be
u=\ep \left (1-\frac{(1-\delta_{2n+2})}{(2n+2)!}\ep^{2n} +O(\ep^{2n+2}) \right ).
\la{ufin}
\ee

Now the partition function at $N$ discrete time step is
\ba
Z_N&=&\int dx \frac1{\sqrt{2\pi \ka_N}}\e^{-\mu_N x^2}=\frac1{\sqrt{2\pi \ka_N}}
\sqrt{\frac{\pi}{\mu_N}}\nn\\
&=&\frac1{\sqrt{2\ka_N\mu_N}}
=\frac1{\sqrt{2(\za_N-1)}}=\frac1{2\sinh(Nu/2)}.
\la{trce}
\ea
The partition function is therefore also
universal and depends on individual short-time propagators
only through the portal parameter $u$ via (\ref{xdef}). This was a very surprising finding
noted by Wang {\it et al.}\cite{wan22}, after integrating various short-time propagators.
It is unsurprising here because we have already shown that the discrete propagator itself is universal.

The $N$-bead thermodynamics energy is
\ba
E_N^T&=& -\frac1{Z_N}\frac{d Z_N}{d \tau}=\frac{du}{d\ep}E_N
\la{zn}
\ea
where 
\be
E_N=\frac12\coth(Nu/2)=\frac12+\frac1{\e^{Nu}-1}
\la{enu}
\ee
is the universal discrete energy. For propagator with $u$ given by (\ref{uep}),
in the convergent limit,
\ba
\lim_{N\rightarrow\infty}E_N
&=&(1+\frac{\tau}{\sinh(\tau)}C\ep^{2n})E(\tau)+\cdots
\la{errde}
\ea
where $E(\tau)$ is the exact free energy
\be
E(\tau)=\frac12\coth(\tau/2).
\ee
The thermodynamics energy therefore converges to leading order in $\ep$ as
\ba
\lim_{N\rightarrow\infty} E^T_N(\tau)
&\rightarrow&\frac{du}{d\ep}(1+\frac{\tau}{\sinh(\tau)}C\ep^{2n})E(\tau)+\cdots \nn\\
&=&(1-(2n+1)C\ep^{2n})(1+\frac{\tau}{\sinh(\tau)}C\ep^{2n})E(\tau)
+\cdots\nn\\
&=&E(\tau)-E(\tau)\left(2n+1-\frac{\tau}{\sinh(\tau)}\right)C\ep^{2n}+\cdots.
\la{eer}
\ea
This shows that if $\za_1$ is {\it correct} to order $2n$, the thermodynamic energy has {\it error} of order $2n$.
Since $\tau/\sinh(\tau)\leq 1$ and goes rapidly to zero with increasing $\tau$, 
at any reasonably large $\tau$,  the energy error is almost entirely 
due to the convergent error of the portal parameter $u$:
\be
E^T_N(\tau)
\rightarrow E(\tau)\left (1-(2n+1)\frac{(1-\delta_{2n+2})}{(2n+2)!}\ep^{2n}+\cdots\right ).
\la{econg}
\ee
This result shows that for $\delta_{2n+2}\leq 1$, the thermodynamic energy 
converges from below, in accordance with the Golden-Thompson inequality\cite{gol65,tho65}. 
Compare (\ref{econg}) to (\ref{ufin}) shows that the convergent error of
the thermodynamic energy is not only of the same order as
the portal parameter, but is $(2n+1)$ times larger. In the case of PA, $n=1$, $\delta_4=0$ 
and one can read off from (\ref{econg})
\be
E^T_N(\tau)
\rightarrow E(\tau)\left (1-\frac1{8}\ep^{2}+\cdots\right ).
\la{epa}
\ee

Thus for the harmonic oscillator,
any short-time propagator can be optimized by simply forcing its coefficient $\za_1(\ep)$
to match the expansion of $\cosh(\ep)$ to maximal order. There is no need to compute the energy separately.

For example, consider a modification of the PA propagator by adding
a double commutator term with parameter $\al$ to the potential operator via
\be
{\cal T}_{TI}(\ep)=
\e^{-\ep\hat V/2-\al\ep^3[\hat V,[\hat T,\hat V]]}
\e^{-\ep \hat T }
\e^{-\ep\hat V/2-\al\ep^3[\hat V,[\hat T,\hat V]]}.
\label{tiop}
\ee
For the harmonic oscillator, $[\hat V,[\hat T,\hat V]]=[V^\prime(x)]^2=x^2$, 
this propagator has 
\be 
\ka_1(\dt)=\dt\quad{\rm and}\quad \mu_1(\dt)=\dt/2+2\al\ep^3,
\la{paia}
\ee
and therefore
\be
\za_1=1+\ka_1\mu_1=1+\frac12 \ep^2+2\al\ep^4.
\la{expza}
\ee
This will match the expansion (\ref{chexp}) to fourth-order in $\ep$ for $\al=1/48$. 
The resulting ${\cal T}_{TI}$ is the correctable\cite{chin04} fourth-order 
Takahashi–Imada (TI) propagator\cite{tak84}, and this is its simplest derivation.
Moreover, the thermodynamics energy now converges, as given by (\ref{econg}) with $n=2$, $\delta_{6}=0$, 
\be
E^T_N(\tau)\rightarrow E(\tau)\left (1-\frac1{144} \ep^{4}+\cdots\right ).
\la{eti}
\ee

This one-line derivation of the TI propagator
illustrates the power of the matching condition (\ref{zacos}) and will now be applied to
higher order algorithms below.


\section {optimizing fourth-order propagators}
\la{opfour}

The complexity of a short-time propagator is measured by the number of its $\hT$ operators.
At the two-$\hT$ level, one can improve the correctable TI algorithm to the truly fourth-order 4A
short-time propagator\cite{chin02} with double commutators distributed by $\al$: 
\be
{\cal T}_{4A}= 
\e^{-\frac16\ep\hV_0}
\e^{-\frac{1}2\ep \hat T}
\e^{-\frac23\ep\hV_1}
\e^{-\frac{1}2\ep \hat T}
\e^{-\frac16\ep\hV_0}
\ee
\ba
&&\qquad\frac16\hV_0=\frac16 \hV+\frac{\al}{2}\frac{\epsilon^2}{72}[\hV,[\hT,\hV]]\nn\\
&&\qquad\frac23\hV_1=\frac23 \hV+(1-\al)\frac{\epsilon^2}{72}[\hV,[\hT,\hV]].
\label{vtbd}
\ea
Applying the two-$\hT$ contraction (\ref{kauv}) with $a=c=\ep/2$ and 
\be
b=\frac23\ep+2(1-\al)\frac{\ep^3}{72},
\ee
immediately gives
\ba
\ka_1&=&a(2+ab)=\ep(1+\frac{\ep}4 b)=\ep(1+\frac16\ep^2+\frac{1-\al}{144}\ep^4)\la{ka4a}\\
\za_1&=&1+\ka_1\left [\frac{ba}{\ka_1}+\ep(\frac16+\al\frac{\ep^2}{72})\right ]\nn\\
&=&1+\frac{\ep^2}{2}+\frac{\ep^4}{4!}+\frac{1+\alpha}{864}\ep^6+\frac{\alpha(1-\alpha)}{10368}\ep^8,
\la{za4a}
\ea
verifying that this is a fourth-order algorithm for all values of $\al$.
Notice that the denominator $\ka_1$ in $\mu_1$ is canceled when when computing their
product and $\za_1$ is only a polynomial in $\ep$. As we will see, this is always the case.
The original 4A algorithm is defined by $\al=0$.  
The choice of $\al=1/5$ forcing
$(1+\alpha)/864=1/6!$ would yield a six-order algorithm with
\be
\za_1(\ep)=1+\frac{\ep^2}{2}+\frac{\ep^4}{24}+\frac{\ep^6}{720}+\frac{\ep^8}{64800}.
\ee
Here, as in the TI case, the optimal distribution of $[\hV,[\hT,\hV]]$  
improves the propagator's order of convergence by two. We will refer to 
this case as the 4A$'$ algorithm.

According to (\ref{econg}), algorithms of the same order $2n$ have 
energy error proportional to 
\be
e_{2n+2}=1-\delta_{2n+2}.
\ee
For discussing sixth-order algorithm in this Section, it is only necessary to
compare their error coefficient $e_8$.   
For the above algorithm, $\delta_8=8!/64800$ and $e_8=0.3778$.

Here, the portal parameter is defined by $u=\cosh^{-1}(\za_1)$ with $\za_1$ given by (\ref{za4a}).
The discrete propagator is given by $\ka_N$ and $\mu_N$ as defined in Sect.\ref{mpro}. This propagator seemingly disagrees with the one given by Kamibayashi and Miura\cite{kam16}. 
Their propagator at $N=1$ looked nothing like the short-time propagator as defined by $\ka_1$ (\ref{ka4a}) and $\za_1$ (\ref{za4a}) above. Surprisingly, this turns out to be a non-trivial
example of ``reparametrization equivalence". In the Appendix, using our universal propagator, 
we can easily derive their density matrix and demonstrate their equivalence in computing all physical observables.

\begin{figure}[hbt]
	\includegraphics[width=0.70\textwidth]{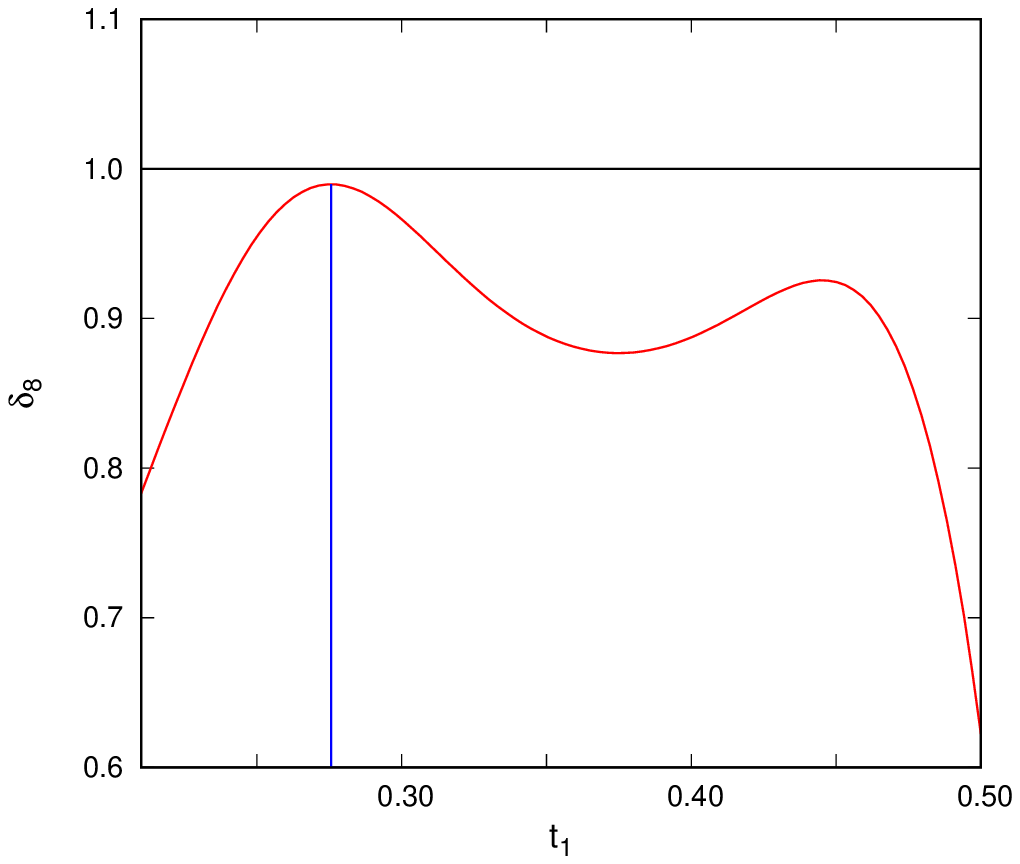}	
	\caption{The coefficient $\delta_8$ as a function of $t_1$
		for the BDA propagator (\ref{algbda}). The vertical line at $t_1=0.27564$ gives
		$\delta_8=0.98967$.
	}
	\la{e8}
\end{figure}

At the three-$\hT$ level, one has the
BDA family of fourth-order short-time propagator\cite{chin02}
with distributed commutators 
given by
\begin{equation}
	{\cal T}_{BDA}=
	{\rm e}^{-v_0\epsilon \hV_0}
	{\rm e}^{-t_1\epsilon \hT} 
	{\rm e}^{-v_1\epsilon \hV_1}
	{\rm e}^{-t_2\epsilon \hT} 
	{\rm e}^{-v_1\epsilon \hV_1}
	{\rm e}^{-t_1\epsilon \hT}
	{\rm e}^{-v_0\epsilon \hV_0},
	\label{algbda}
\end{equation}
\be
t_2=1-2t_1,\quad
v_1=\frac1{12t_1(1-t_1)},\quad 
v_0=\frac12-v_1,
\ee
\ba
v_0\hV_0&=&v_0\hV+\al u_0\epsilon^2[\hV,[\hT,\hV]],\nn\\
v_1\hV_1&=&v_1\hV+(1-\al)u_0\epsilon^2[\hV,[\hT,\hV]],\nn\\
u_0&=&{1\over 48}\biggl[{1\over{6t_1(1-t_1)^2}}-1\biggr],
\label{vtbd}
\ea
The two free parameters are ${1\over 2}(1-{1\over{\sqrt 3}})\leq t_1\leq {1\over 2}$ and $0\leq\al\leq 1$.

The three-operator contraction (\ref{threet}) 
is directly applicable with
\be
a=t_1\ep,\quad b=v_1\ep+2(1-\al)u_0\ep^3,\quad c=t_2\ep
\ee
yielding immediately
\ba
\ka_1&=&(1+ab)(2a+abc+c),\quad\mu_1=\frac{b}{1+ab}+v_0\ep+2\al u_0\ep^3\nn\\
\za_1&=&1+\ka_1\mu_1=1+(2a+abc+c)[b+(1+ab)(v_0\ep+2\al u_0\ep^3)]\la{zabb}\\
&=&1+\frac{\ep^2}{2!}+\frac{\ep^4}{4!}+\delta_6(\al,t_1)\frac{\ep^6}{6!}
+\delta_8(\al,t_1)\frac{\ep^8}{8!}+\cdots.
\ea
Again, (\ref{zabb}) shows that $\za_1$ is only a polynomial function of $\ep$.
Since the coefficients are correct up to $\ep^4$, this is a fourth-order propagator
for all values of $\al$ and $t_1$.
Solving $\delta_6(\al,t_1)=1$ for $\al$,
\be
\al=\frac{5-78 t_1+474 t_1^2-1404 t_1^3+2088 t_1^4-1440 t_1^5+360 t_1^6}
{10 (1-6t_1+12 t_1^2-6 t_1^3)^2},
\ee
yields a sixth-order propagator. The optimal $t_1$ can then be determined by 
seeing where $\delta_8(\al(t_1),t_1)$ comes closest to one.
This is shown in Fig.\ref{e8} with $t_1$ at $0.27564$, corresponding to $\al=0.171438$,
with $\delta_8=0.98967$. This gives an
energy error $e_8=0.0103$ which is $\approx 30$ times smaller 
than the 4A$'$ case and is close to being an eight-order algorithm.

At the four-$\hT$ level, Sakkos {\it et al.} \cite{sak09} have done extensive PIMC simulations 
using the ACB form of the propagator\cite{chin02,san05} with distributed commutators: 
\be
{\cal T}_{ACB}\equiv
  {\rm e}^{ -t_0\ep\hT}
  {\rm e}^{ -v_1\ep\hV_1}
  {\rm e}^{ -t_1\ep\hT}
  {\rm e}^{ -v_2\ep\hV_2}
  {\rm e}^{ -t_1\ep\hT}
  {\rm e}^{ -v_1\ep\hV_1}
  {\rm e}^{- t_0\ep\hT}\, , \la{algacop}
\ee
where 
\be
t_1={1\over 2}-t_0 ,\quad
v_2=1-2v_1,\quad
v_1={1\over 6}{1\over{(1-2 t_0)^2}} 
\la{acofac}
\ee
\ba
v_1\hV_1&=&v_1\hV+\frac{\al}2 u_0 \ep^2 [\hV,[\hT,\hV]]\, , \nn\\
v_2\hV_2&=&v_2\hV+(1-\al)u_0\ep^2[\hV,[\hT,\hV]]\, , \la{vtac2}\\
u_0&=&{1\over 12}\biggl[1-{1\over{1-2t_0}}+{1\over{6(1-2t_0)^3}}\biggr].\la{uo}
\ea

\begin{figure}[hbt]
	\includegraphics[width=0.70\textwidth]{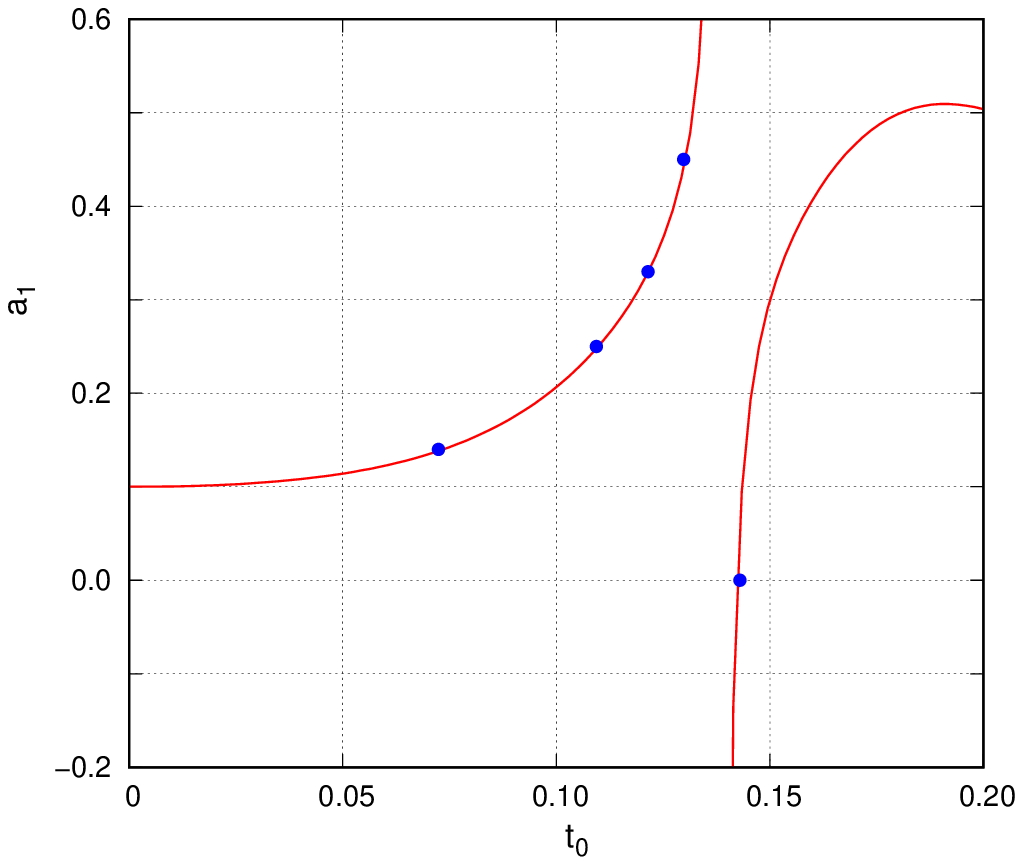}	
	\caption{ The five optimal values of $(t_0,a_1)=$ 
		(0.0724,0.14),(0.1094,0.25),(0.1215,0.33),(0.1298,0.45) and (0.1430,0) 
		found empirically by Sakkos {\it et al.}\cite{sak09} for the ACB propagator (\ref{algacop})
		plotted against $a_1=\al/2$ with $\al$ given by (\ref{alp}).
	}
	\la{sak}
\end{figure}

\begin{figure}[hbt]
	\includegraphics[width=0.70\textwidth]{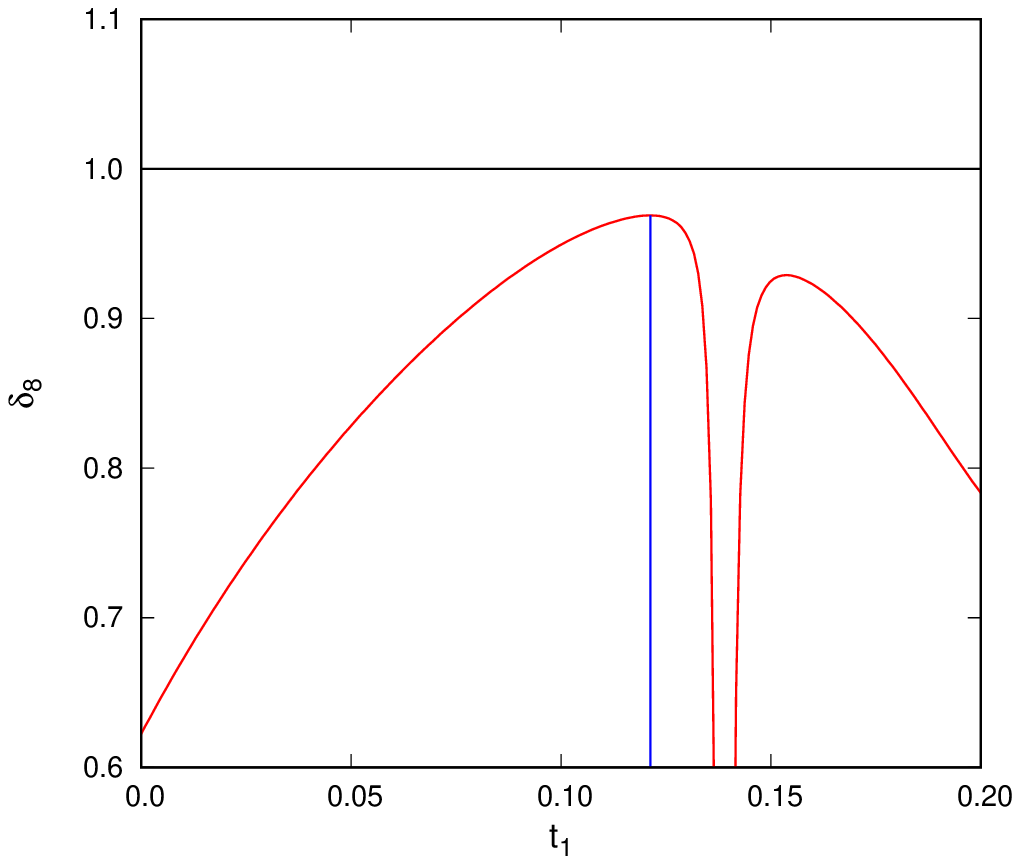}	
	\caption{ The coefficient $e_8$ as a function of $t_1$
		for the ACB propagator (\ref{algacop}). 
		The vertical line at $t_1=0.1213$ gives
		$\delta_8=0.9689$.
	}
	\la{e8zak}
\end{figure}

The central two $\hT$ in (\ref{algacop}) can be contracted with
\be
a=c=t_1\ep\quad{\rm and}\quad b=v_2\ep+2(1-\al)u_0\ep^3,
\la{abc}
\ee
yielding
\be
{\cal T}_{ACB}=
  {\rm e}^{ -t_0\ep\hT}
  {\rm e}^{ -v_1\ep\hV_1}
  {\rm e}^{ -\mu\hV}
  {\rm e}^{ -\ka\hT}
  {\rm e}^{ -\mu\hV}
  {\rm e}^{ -v_1\ep\hV_1}
  {\rm e}^{- t_0\ep\hT}
\ee
where, again from (\ref{kauv}),
\be
 \ka=a(2+ab),\quad\mu=\frac{b}{2+a b}.
 \la{kanu}
\ee
The remaining three-$\hT$ form can be contracted according to (\ref{threet}) 
(with $a'$, $b'$, $c'$ in places of  $a$, $b$, $c$),
\be
a'=t_0\ep,\quad b'=\mu+v_1\dt+\al u_0\dt^3\quad c'=\ka
\la{abcp}
\ee
yielding finally,
\be
\ka_1=(1+a'b')(2a'+c'+a'b'c'),\qquad\mu_1=\frac{b'}{(1+a'b')}.
\la{ka1}
\ee
Again, $\za_1$ is a polynomial function of $\ep$ with expansion
\ba
\za_1&=&1+\ka_1\mu_1=1+b'(2a'+c'+a'b'c')\nn\\
&=&1+\frac{\ep^2}{2}+\frac{\ep^4}{24}+
+\delta_6(\al,t_1)\frac{\ep^6}{6!}
+\delta_8(\al,t_1)\frac{\ep^8}{8!} +\cdots.
\la{zabc}
\ea
Solving $\delta_6(\al,t_0)=1$ for $\al$,
\be
\al=\frac{1-18t_0+144t_0^2-552t_0^3+576t_0^4}{5-90t_0+540t_0^2-840t_0^3-2880t_0^4+8640t_0^5-5760t_0^6}
\la{alp}
\ee
again produces a sixth-order algorithm. This same equation for $\al$ has been previously obtained in a
much more elaborate study\cite{san05} of the classical harmonic oscillator.

In Fig.\ref{sak}, the parameter $a_1=\al/2$, used by Sakkos {\it et al.}\cite{sak09}, is plotted
against the five optimal values of ($t_0$,$a_1$) they found by trial and error. The agreements are
excellent. Moreover, their Fig.4 shows that the best among the five is at $t_0=0.1215$
with $a_1=0.33$.
This can now be understood as the maximization of $\delta_8(\al(t_0),t_0)=0.9689$
at $t_0=0.1213$, as shown in Fig.\ref{e8zak}. This corresponds to $\al=0.6553$, with 
error coefficient $e_8=0.0311$, three times larger than the BDA case. 

\begin{figure}[hbt]
	\includegraphics[width=0.70\textwidth]{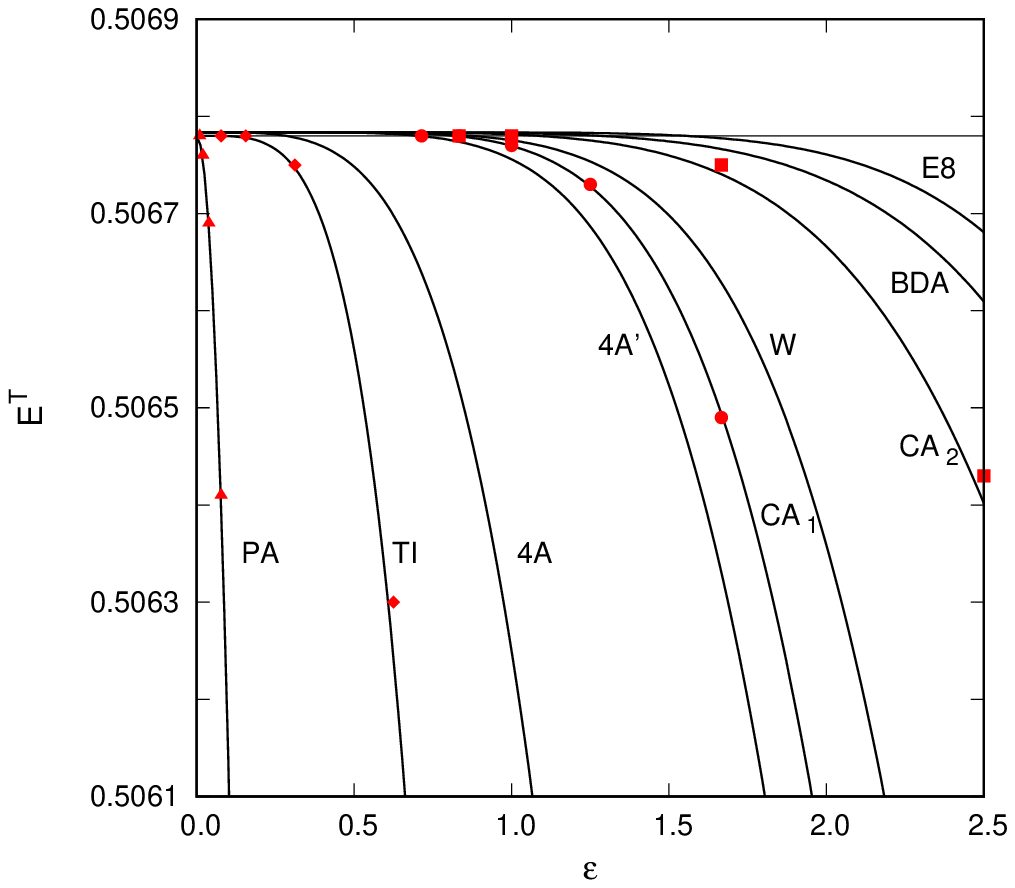}	
	\caption{ Plotting symbols are Sakkos {\it et al.}'s PIMC data\cite{sak09} for four short-time
	propagators PA, TI, CA$_1$ and CA$_2$.
	Solid lines are analytical thermodynamic energies for all short-time propagators discussed in this work. 
	See text for details.
	}
	\la{congall}
\end{figure}

In Fig.\ref{congall}, PIMC results from Sakkos {\it et al.}'s Table I are plotted for
PA, TI and two cases of ACB which they designated as CA$_1$ ($a_1=0,t_0=0.1430$) and CA$_2$ ($a_1=0.33,t_0=0.1215$).
For PA, TI, their data are well described by the leading order thermodynamic energy (\ref{epa}) and (\ref{eti}) at $\tau=5$. Such a leading order description is clearly inadequate at $\ep>1$. Fortunately, the
thermodynamic energy can be exactly given for any short-time propagator solely in terms of the
product coefficient $\za_1(\ep)$. Given $\za_1(\ep)$, $u(\ep)$ 
is known from its fundamental definition (\ref{udef}) and hence
\be
\frac{du}{d\ep}=\frac1{\sqrt{\za_1^2-1}}\frac{d\za_1}{d\ep}.
\la{dude}
\ee
Since $\za_1(\ep)$ is a polynomial in $\ep$, its derivative is known analytically.
The discrete thermodynamic energy from (\ref{zn}) is then completely defined by 
\be
E_N^T=\frac1{\sqrt{\za_1^2-1}}\frac{d\za_1}{d\ep}\left (\frac12+\frac1{\e^{N\ep(u/\ep)}-1}\right).
\la{znan}
\ee
If $\ep$ is given initially, then $E_N^T$ is the energy at increasingly larger discrete
time $\tau_N=N\ep$. However, if $\tau$ $(=N\ep)$ is fixed, then (\ref{znan}) gives the convergent
energy as a function of $\ep$:
\be
E^T(\ep)=\frac1{\sqrt{\za_1^2-1}}\frac{d\za_1}{d\ep}\left (\frac12+\frac1{\e^{\tau(u/\ep)}-1}\right).
\la{zntau}
\ee
In Fig.\ref{congall} we plot the above energy for all sixth-order short-time propagators discussed in this Section
at $\tau=5$. The discrete energy inside the parenthesis is a slow varying function of $\ep$ because
$(u/\ep)\approx 1$. Most of the variation is due to the prefactor (\ref{dude}). One now sees excellent 
agreements with Sakkos {\it et al.}'s CA$_1$ and CA$_2$ results. W is Wang {\it et al.}'s
sixth-order algorithm\cite{wan22} and E8 is a new eight-order propagator. 
Both will now be described in the following Section.

\section {An eight-order propagators}
\la{eighto}

\begin{figure}[hbt]
	\includegraphics[width=0.70\textwidth]{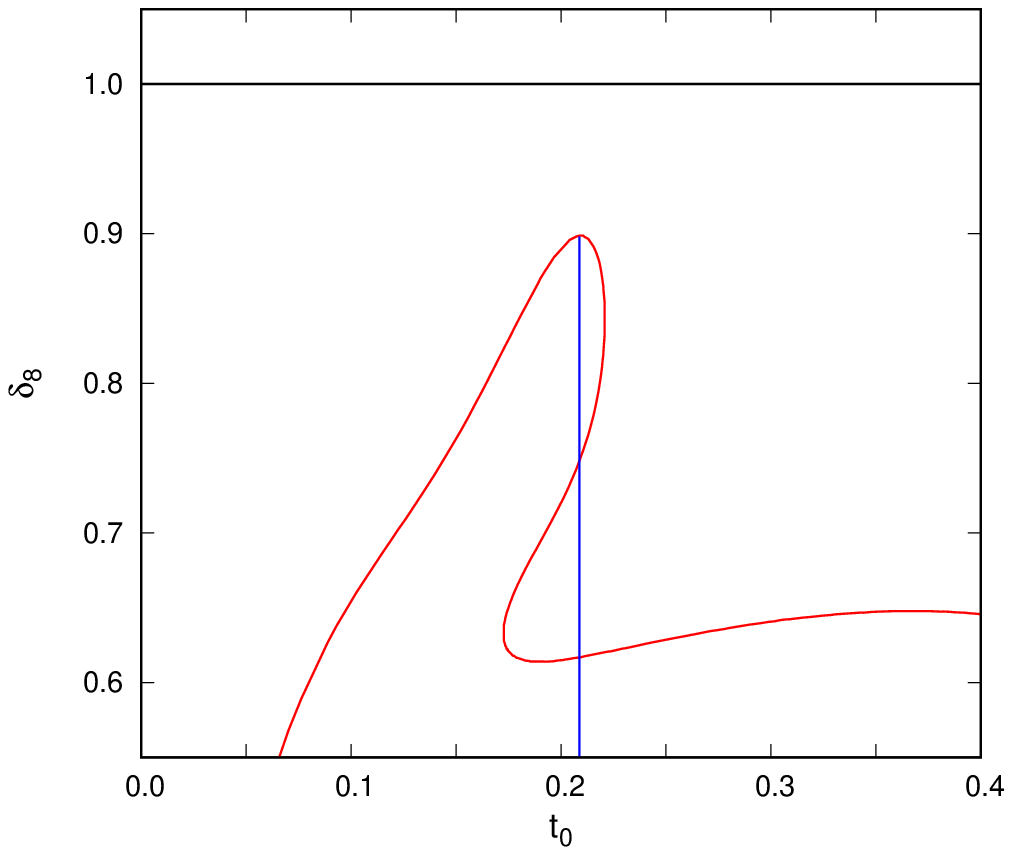}	
	\caption{ The coefficient $\delta_8$ as a function of $t_0$
		for for Wang {\it et al.}'s\cite{wan22} sixth-order
		propagator (\ref{wang}) with $c_0=0$. 
		The vertical line at $t_0=0.209$ gives $\delta_8=0.8987$.
		}
	\la{e8noc}
\end{figure}

\begin{figure}[hbt]
	\includegraphics[width=0.70\textwidth]{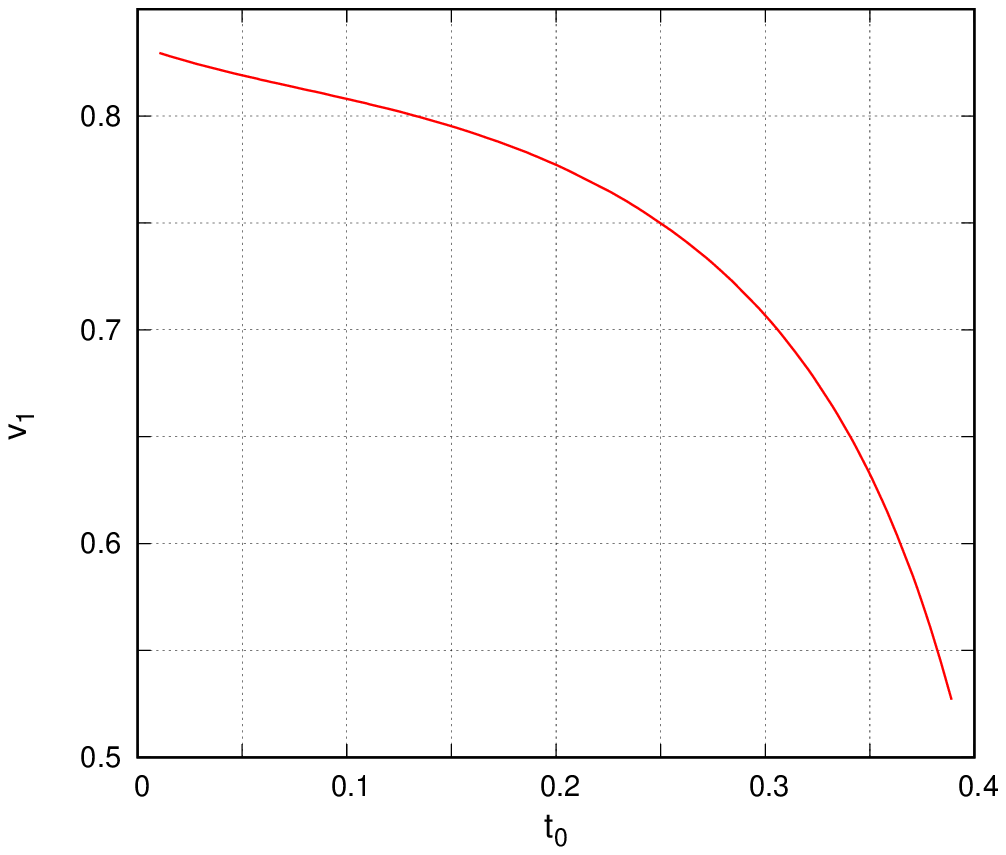}	
	\caption{ The parameter $v_1$ as a function of $t_0$ for yielding an eight-order algorithm
		satisfying $\delta_8(c_0(t_0,v_1),t_0,v_1)=1$ for
		propagator (\ref{wang}).
	}
	\la{e10}
\end{figure}

The previous two cases are true fourth-order propagators with $v_1$ fixed to
remove the unwanted commutator $[\hT,[\hT,\hV]]$. The
two free parameters $\al$ and $t_0$ (or $t_1$) can then be 
optimized to achieve six-order, but not quite eight-order, convergence. 
Recently, Wang {\it et al.}\cite{wan22} suggested that one should simply require the
fourth-order propagator to be correctable, like TI, so that $v_1$ can be freed for optimization.
This can done by choosing $u_0$, as in (\ref{uone}) below, so that the coefficient of $[\hT,[\hT,\hV]]$
matches the coefficient of $[\hV,[\hT,\hV]]$ in the error Hamiltonian\cite{chin04}.
Their $g4T3V$ propagator is of the same form as (\ref{algacop}), but we will follow their notations closely:
\be
{\cal T}_{g4T3V}\equiv
{\rm e}^{ -t_0\ep\hT}
{\rm e}^{ -v_0\ep\hV_0}
{\rm e}^{ -t_1\ep\hT}
{\rm e}^{ -v_1\ep\hV_1}
{\rm e}^{ -t_1\ep\hT}
{\rm e}^{ -v_0\ep\hV_0}
{\rm e}^{- t_0\ep\hT},
\la{wang}
\ee
\ba
t_1&=&{1\over 2}-t_0,\quad v_0=\frac{1-v_1}2\nn\\
v_0\hV_0&=&v_0\hV+c_0 u_0\ep^2[\hV,[\hT,\hV]],\nn\\
v_1\hV_1&=&v_1\hV+(1-2c_0)u_0\ep^2\,[\hV,[\hT,\hV]],\nn\\
u_0&=&{1\over 24}\left[ (12t_0^2-1)(1-v_1)+(2-6t_0)(1-v_1)^2+v_1^2  \right].
\la{uone}
\ea
The crucial difference here is that $0\leq v_1\leq 1$ is
no longer given by (\ref{acofac}), but is a free parameter along with  
$0\leq t_0\leq 1/2$ and $0\leq c_0\leq 1/2$.

The same sequence of operator contractions (\ref{abc}) to (\ref{zabc}) as in the last case,
but with 
\be
a=c=t_1\ep\quad{\rm and}\quad b=v_1\ep+2(1-2c_0)u_0\ep^3,
\ee
\be
a'=t_0\ep,\quad b'=\frac{b}{2+a b}+v_0\dt+2c_0 u_0\dt^3,\quad c'=a(2+ab)
\ee
now gives
\ba
\za_1&=&1+\ka_1\mu_1=1+b'(2a'+c'+a'b'c')\nn\\
&=&1+\frac{\ep^2}{2}+\frac{\ep^4}{4!}
+\delta_6(c_0,t_0,v_1)\frac{\ep^6}{6!}
+\delta_8(c_0,t_0,v_1)\frac{\ep^8}{8!} +\delta_{10}(c_0,t_0,v_1)\frac{\ep^{10}}{10!}+\cdots.
\ea

Wang {\it et al.}\cite{wan22} set $c_0=0$ to minimize the evaluation of $[\hV,[\hT,\hV]]$.
In this case, solving $\delta_{6}(t_0,v_1)=1$ 
gives the same function $v_1(t_0)$ as in their Fig.2.
Plotting the resulting $\delta_{8}(t_0,v_1(t_0))$, as done in Fig.\ref{e8noc}, determined that
the optimal sixth-order propagator is at $t_0=0.209$, in agreement with Wang {\it et al.}'s\cite{wan22}
stated result of $t_0=0.209114$. The resulting $\delta_8=0.8987$ gives $e_8=0.1013$,
which is smaller than 4A$'$ ($e_8$=0.38) and CA$_1$ ($e_8$=0.25) but is 
an order of magnitude larger than allowing $c_0$ $(=\al/2)$ to vary in both ACB and BDA.
Its energy convergent curve is plotted as W in Fig.\ref{congall}.

\begin{figure}[hbt]
	\includegraphics[width=0.70\textwidth]{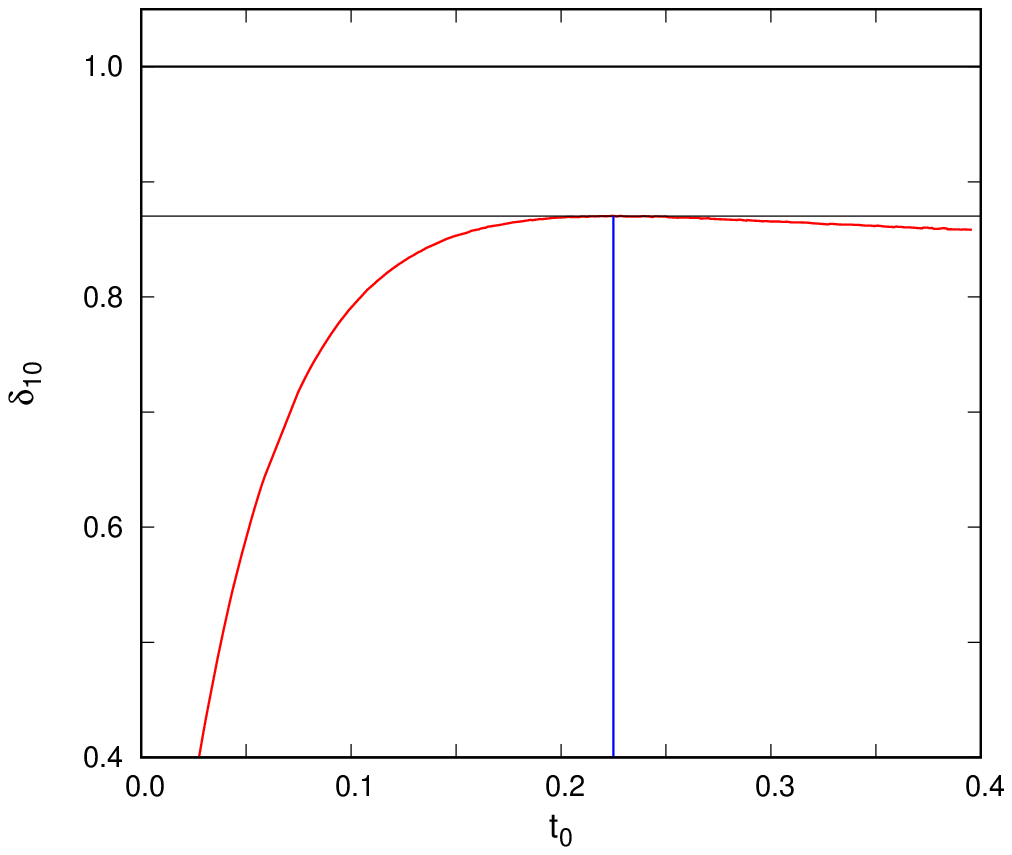}	
	\caption{ The coefficient $\delta_{10}$ as a function of $t_0$ for
		propagator (\ref{wang}). The vertical line at $t_0=0.2257$ gives
		$\delta_{10}=0.8702$.
	}
	\la{e10opt}
\end{figure}

However, allowing $c_0$ to vary can now improve the convergence from six to
eight order.
Solving $\delta_{6}(c_0,t_0,v_1)=1$ for $c_0(t_0,v_1)$ yields a sixth-order algorithm as before. Solving
$\delta_{8}(c_0(t_0,v_1),t_0,v_1)=1$ numerically for $v_1$ as a function of $t_0$, as shown in Fig.\ref{e10},
now yields an {\it eight-order} propagator. 

In Fig.\ref{e10opt}, $\delta_{10}(c_0(t_0,v_1(t_0)),t_0,v_1(t_0))=0.8702$ 
at $t_0=0.2257$, gives the optimal eight-order algorithm with
$v_1=0.7646$, $c_0=0.02976$ and $\delta_{10}=0.13$. Its energy convergent curve is plotted
as E8 in Fig.\ref{congall}, better than both CA$_2$ (the optimal version of ACB) and BDA.
PA's energy near $0.5067$, at $\ep\approx 0.04$, is matched by E8's energy near $\ep\approx2.5$, at a time
step $\approx 60$ times larger.

\section {Conclusions and future directions}
\la{con}

This work has shown that, on the basis of a single operator contraction identity,
everything about the harmonic oscillator path integral can be known simply,
from deriving its universal discrete propagator, optimizing all short-time propagators, to 
the exact formula for the thermodynamic energy.
The central role is played by the product coefficient $\za_1(\ep)$, out of which the
portal parameter $u$ is derived, the discrete propagator defined, and whose derivative and
series expansion determined the thermodynamic energy and sequential steps of optimization. Because of
this detailed knowledge, all previous published results can now be understood analytically 
and a new eight-order algorithm derived.  

   A natural follow up to this work would be a detailed study on the convergence of
the thermodynamic and Hamiltonian ground state energy as a function of $N$. Since the harmonic oscillator is 
separable in any dimension, the discrete propagator derived here is applicable to harmonic oscillators in any dimension. In particular, it will be of interest to study the anti-symmetrized discrete fermion propagator 
in two or three dimensions, to gain further insights on how to ameliorate the sign problem\cite{chin15}
in quantum dots using higher order propagators.

\bigskip\bigskip\bigskip

\noindent
{\bf Acknowledgment }

The author would like to thank Fernando Casas and his colleague, for suggesting that
the universal coefficients may be related to Chebyshev polynomials.

\appendix*
\section {Deriving Kamibayashi and Miura's 4A density matrix }

Kamibayashi and Miura\cite{kam16} stated their density matrix for the 4A propagator
in their Eq.(15) as
\be
\rho_{FOD}(x,x';\beta)
\propto\exp\left[-\sqrt{\frac{r}{r'}}\frac{\sinh\varphi}{\ddt}\frac1{2\sinh(\beta\varphi/\ddt)}
\{(x^2+x'^2)\cosh(\beta\varphi/\ddt)-2xx'\} \right],
\la{fod}
\ee
where (we have set their $\ga=\al$)
\ba
r&=&1+\frac13\ddt^2+\frac{\al}9 \ddt^4,\nn\\
r'&=&1+\frac23\ddt^2+\frac19 (1-\al)\ddt^4.
\ea
At $\beta=\ddt$, (\ref{fod}) would give,
\be
\rho_{FOD}(x,x';\ddt)
\propto\exp\left[-\sqrt{\frac{r}{r'}}\frac{1}{2\ddt}
\{(x^2+x'^2)\sqrt{r r'}-2xx'\} \right],
\la{fod2}
\ee
bearing no resemblance to $\ka_1$ (\ref{ka4a}) and $\za_1$ (\ref{za4a}).
However, Kamibayashi and Miura's short-time operator, their Eq.(5), {\it starts out} at $\beta=2\ddt$:
\be
\e^{-2\ddt\hH}=\e^{-\frac13\ddt \hV_e}\e^{-\ddt \hT}\e^{-\frac43 \ddt \hV_m}
\e^{-\ddt \hT}\e^{-\frac13\ddt \hV_e}.
\la{eq5}
\ee 
This means that
\be
\rho_{FOD}(x,x';2\ddt)=G_1(x,x';\ep=2\ddt).
\ee
Therefore, in terms of $\ddt$, $\ka_1$ (\ref{ka4a}) and $\za_1$ (\ref{za4a}) now read
\be
\ka_1(\ep=2\ddt)=2\ddt (1+\frac23\ddt^2+\frac19 (1-\al)\ddt^4)=(2\ddt) r'
\ee
and
\ba
\za_1
&=&1+\frac{\ep^2}{2}+\frac{\ep^4}{4!}+\frac{1+\alpha}{864}\ep^6
+\frac{\alpha(1-\alpha)}{10368}\ep^8\nn\\
&=&1+2\ddt^2+\frac{2\ddt^4}{3}+\frac{2}{27}(1+\alpha)\ddt^6+\alpha(1-\alpha)\frac{2}{81}\ddt^8\nn\\
&=&1+2(rr'-1)=2rr'-1.
\la{par}
\ea
In this work, we have defined the portal parameter $u$ by
\be
\za_1(\ep)=\cosh(u).
\ee
However, one can also define the portal parameter $\varphi$ via
\be
\za_1(2\ddt)=\cosh(2\varphi)=2\cosh^2(\varphi)-1,
\ee
which then from (\ref{par}) gives
\be
\cosh(\varphi)=\sqrt{rr'}.
\ee
The algorithm advances $2\ddt$ per application and one has similarly
\be
\za_N(2\ddt)=\cosh(N 2\varphi).
\ee
This means that Kamibayashi and Miura's density matrix only holds for $\beta/\ddt=2N$, an even number.
Recalling (\ref{knx}) then gives
\ba
\ka_N&=&\frac{(2\ddt) r'}{\sinh(2\varphi)}\sinh(N2\varphi)\nn\\
&=&\frac{(2\ddt) r'}{2\sinh\varphi\cosh\varphi}\sinh(N2\varphi)\nn\\
&=&\frac{(\ddt) r'}{\sinh\varphi\sqrt{r r'}}\sinh(N2\varphi)\nn\\
&=&\sqrt{\frac{ r'}{r}}\frac{\ddt}{\sinh\varphi}\sinh(N2\varphi)
\la{knpx}
\ea
reproducing (\ref{fod}) when $\beta/\ddt=2N$. However, when computing the partition function  
(\ref{trce}),
\be
Z_N=\frac1{2\sinh(N 2\varphi/2)},
\ee
this restriction no longer matters, because one is simply reparametizing $u$ as $2\varphi$.

\end{document}